\documentclass[journal,10pt]{IEEEtran}

\usepackage{array}
\usepackage{cite}
\usepackage{mathtools}
\usepackage{amssymb}
\usepackage{microtype}
\usepackage[noend]{algpseudocode}
\usepackage{multirow}
\usepackage[hidelinks]{hyperref}
\makeatletter
\renewcommand{\ALG@beginalgorithmic}{\small}
\makeatother
\algnewcommand\HEADER{\item[\textbf{Algorithm}]}
\algnewcommand\INPUT{\item[\textbf{Input:}]}%
\algnewcommand\OUTPUT{\item[\textbf{Output:}]}%
\algnewcommand\algorithmicforeach{\textbf{for each}}
\algdef{S}[FOR]{ForEach}[1]{\algorithmicforeach\ #1\ \algorithmicdo}
\makeatletter
\newcommand{\algrule}[1][.4pt]{\par\vskip.25\baselineskip\hrule height #1\par\vskip.25\baselineskip}
\makeatother

\usepackage{tikz}
\usetikzlibrary{positioning, angles, quotes}
\tikzstyle{every node}=[font=\footnotesize]
\usepackage{pgfplots}
\pgfplotsset{compat=newest}
\pgfplotsset{label style={font=\footnotesize},
tick label style={font=\footnotesize},
xlabel shift = -2 pt,
legend style={font=\footnotesize},
legend cell align={left},
xtick style={color=black, thin},
ytick style={color=black, thin},
tick align=outside,
tick pos=left,
axis on top=true}

\DeclareMathOperator{\diag}{diag}
\DeclareMathOperator{\sign}{sign}
\DeclareMathOperator*{\minimize}{minimize}
\DeclarePairedDelimiter\norm{\lVert}{\rVert}

\newtheorem{definition}{Definition}

\newtheorem{myrule}{Principle}

\title{
Resilient Branching MPC for Multi-Vehicle Traffic Scenarios Using Adversarial Disturbance Sequences
}

\author{Victor Fors, Bj{\"o}rn Olofsson, and Erik Frisk
\thanks{This work was supported by Excellence Center at Link{\"o}ping--Lund in Information Technology (ELLIIT).
This work was partially supported by the Wallenberg AI, Autonomous Systems and
Software Program (WASP) funded by the Knut and Alice Wallenberg Foundation.}
\thanks{Victor Fors and Erik Frisk are with the Division of Vehicular Systems, Department of Electrical Engineering,
        Link{\"o}ping University, SE-581~83 Link{\"o}ping, Sweden.
        (e-mail: victor.fors@liu.se; erik.frisk@liu.se).}%
\thanks{Björn Olofsson is with the Division of Vehicular Systems, Department of Electrical Engineering,
        Link{\"o}ping University, SE-581~83 Link{\"o}ping, Sweden,
        and also with the Department of Automatic Control, Lund University, SE-221~00 Lund, Sweden (e-mail: bjorn.olofsson@liu.se).}%
}

\newcommand\copyrighttext{%
  \footnotesize \textcopyright \color[gray]{0.3} 2022 IEEE. Personal use of this material is permitted.  Permission from IEEE must be obtained for all other uses, in any current or future media, including reprinting/republishing this material for advertising or promotional purposes, creating new collective works, for resale or redistribution to servers or lists, or reuse of any copyrighted component of this work in other works.
  DOI: \href{https://doi.org/10.1109/TIV.2022.3168772}{10.1109/TIV.2022.3168772}}%
\newcommand\copyrightnotice{%
\begin{tikzpicture}[remember picture,overlay]%
\node[anchor=south] at (current page.south) {{\parbox{\dimexpr\textwidth-\fboxsep-\fboxrule\relax}{\copyrighttext}}};
\end{tikzpicture}%
}%

\begin{document}

\maketitle

\begin{abstract}
An approach to resilient planning and control of autonomous vehicles
in multi-vehicle traffic scenarios is proposed. The proposed method is based on model predictive control (MPC), where alternative predictions of the
surrounding traffic are determined automatically such that they are
intentionally adversarial to the ego vehicle. This provides robustness
against the inherent uncertainty in traffic predictions. To reduce
conservatism, an assumption that other agents are of no ill intent is formalized. Simulation results from highway driving scenarios show that the proposed method in real-time negotiates traffic situations out of scope for a nominal MPC approach and performs favorably to state-of-the-art reinforcement-learning approaches without requiring prior training. The results also show that the proposed method performs effectively, with the ability to prune disturbance sequences with a lower risk for the ego vehicle.
\end{abstract}

\section{Introduction}

\IEEEPARstart{A}{utonomous} vehicles have to make the same fast decisions as we humans make on the fly.
Other road occupants could be expected to follow traffic rules and take logical actions, but there is no guarantee that they will meet these expectations.
Reports from autonomous-driving experiments on public roads indicate that many accidents with other vehicles or incidences where a safety driver elects to take control to prevent a potential collision are caused by human error \cite{schwall2020waymo}. While challenging, the potential of autonomous driving in terms of increased safety is high \cite{BOLN:2021:TITS}.

\copyrightnotice%
Traffic prediction is an active research field encompassing a variety of methods \cite{https://doi.org/10.1186/s40648-014-0001-z,9158529}, methods that can be incorporated in a decision-making and planning framework.
Even with good motion or intent predictions, the inherent uncertainty introduced by traffic is large compared to the typical uncertainty stemming from sensor measurements and modeling errors.
With other agents on the road having similar capabilities as the ego vehicle,
the potential deviations from the predicted behavior can have the same magnitude as the control-input limits.
To improve robustness and resilience, and thereby to increase safety
and reduce the number of accidents, the decision-making and planning of an
autonomous vehicle should account for these large, often multi-modal,
uncertainties arising from the surrounding traffic.
While a number of methods in literature address this,
they rely on prior data of the traffic behavior, which may not cover all possible critical events.
This paper demonstrates a method that does not use such knowledge.

\begin{figure}
\centering
\includegraphics{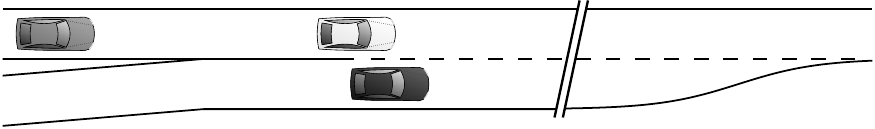}\\
\caption{\label{fig:scenario}%
Highway merging scenario with multiple vehicles interacting.}
\end{figure}

An interesting case study to highlight challenges for such a method is highway merging as illustrated in Fig.~\ref{fig:scenario}.
In this scenario, the bottom vehicle is entering the highway and must merge before the entry lane ends.
There is a trade-off between safety and comfort, because merging faster may be safer while keeping accelerations limited is more comfortable.
The problem is non-convex from a decision point-of-view, with the possibility to merge in between or on either side of the other vehicles.
Which side it is better to merge on can change depending on the actions of the other vehicles, meaning that the best merging maneuver is dependent on the uncertain behavior of these vehicles.
One consequence of this is that a larger disturbance in the traffic is not necessarily harder to handle than a moderate disturbance, as previously occupied space can be freed up.

\subsection{Related Research} \label{sec:related_research}

With good knowledge of the environment, dynamic-programming methods such as lattice planning using A* can be used to efficiently plan a path and trajectory \cite{https://doi.org/10.1002/rob.21908}.
A weakness of the method comes from the computational requirements when the search space is large, which may require a rough discretization of the state space and a limited set of control-input segments.
A flexible method for planning is Monte-Carlo tree search (MCTS) \cite{6145622}, which has been applied to the domain of motion planning for vehicles \cite{7535424}.
Similar to dynamic-programming methods, the computational requirements of Monte-Carlo tree search in a continuous domain such as the one of autonomous driving increase with a finer discretization of the action space.

Another approach is reinforcement-learning methods such as Deep Q-Network (DQN) \cite{article}, where a neural network learns a control policy from training on a large number of training episodes. 
Originally demonstrating its prowess in Atari games, it has also been applied to the domain of motion planning for vehicles \cite{8569568}.
Combining MCTS with neural network policies has found success in board games such as Go \cite{alphago} and has also been applied to the domain of motion planning for vehicles \cite{8911507}.
The DQN method has been further developed for increased performance into methods such as
Dueling DQN \cite{10.5555/3045390.3045601} and Quantile Regression DQN (QR-DQN) \cite{Dabney_Rowland_Bellemare_Munos_2018}.
The DQN methods only work with discrete action spaces. Examples of reinforcement-learning methods that can handle continuous action spaces are Proximal Policy Optimization (PPO) \cite{schulman2017proximal} and Truncated Quantile Critics (TQC) \cite{pmlr-v119-kuznetsov20a}.
A drawback with reinforcement-learning methods is that the learned policy can have difficulty in situations not covered by the training episodes.

A class of optimization methods that handle large state and action spaces well are the direct methods for optimal control \cite{Diehl}, where the continuous-time state and input are approximated, e.g., using polynomials, and the resulting finite-dimensional optimization problem is solved using numerical methods.
A disadvantage of these methods is that only a locally optimal solution is obtained, which poses a problem when considering non-convex problems.
A way to mitigate this is by combining direct optimal control with a global optimization method such as lattice planning using A* \cite{9084267}.
An alternative method of mitigation is based on identifying
collision-free driving corridors based on reachability analysis
\cite{9170864,9434948} or convex optimization \cite{IV_Morsali_2021}.
In Model Predictive Control (MPC), an open-loop optimization problem is typically solved using a receding horizon where only the first computed sample of the control input in the prediction horizon is applied to the system and the remaining samples are discarded, and then the process is repeated with an updated optimization problem for the new initial condition at the next sample \cite{MORARI1999667}.

A problem with computing an open-loop prediction is that uncertainties in the system model are not explicitly dealt with.
There are many variations to the traditional MPC formulation (from now on referred to as nominal MPC) to handle these uncertainties.
The robust open-loop MPC formulation \cite{KOTHARE19961361} computes a control sequence that is feasible for all possible disturbances, often formulated as a minimax problem.
This open-loop accounting of disturbances results in conservative solutions and can turn a feasible problem infeasible.
Stochastic MPC \cite{9410387} uses statistical distributions to model the uncertainty, transforming the robust hard constraints into chance constraints, allowing some probability of constraint violation.
While the formulation enables less conservative planning, constraints have a certain probability to be violated, with less likely events being more likely to result in constraint violation.
In tube MPC, the problem is reformulated as finding a sequence of control laws \cite{LANGSON2004125,lopez2019dynamic}.
A general feedback policy is prohibitively difficult to compute online, so the solution is approximated as a feedback policy that keeps the system in an invariant tube around the desired trajectory.
Forming these invariant tubes poses a challenge for the domain of motion planning in multi-vehicle scenarios, because the desired driving corridor is dependent on the uncertain behavior of the other vehicles.

The notion of feedback present in the prediction can also be included as in feedback MPC \cite{704989}, 
where each disturbance realization results in a branching of the predicted trajectory, forming a scenario tree of predicted inputs and states (further described in Sec.~\ref{sec:principle_of_planner}).
The resulting problem size grows exponentially with the prediction horizon $N$ and is thus in practice only possible to solve for very short prediction horizons.
Scenario MPC \cite{7990647,9133136} addresses this by sampling a limited number of disturbance realization from a statistical model of the uncertainty.
The result is similar to stochastic MPC in that less likely events are less likely to be accounted for.
On the opposite end, contingency MPC \cite{6497657,alsterda2021contingency} considers only an identified hazardous event,
where the resulting alternative trajectory corresponding to the hazardous event occurring is selectively feasible to the identified hazard.
A challenge lays in how to identify each such hazard.

\subsection{Contributions}

The main contribution is a motion planner and controller called
Adversarial Disturbance-Sequence Branching MPC, denoted ADSB-MPC for
short.  In the domain of multi-vehicle traffic scenarios, ADSB-MPC accounts for uncertainty introduced by traffic and
does this without relying on a statistical distribution of
possible scenarios as in Scenario MPC, identified hazardous events as
in Contingency MPC, or multiple training episodes in related scenarios as in reinforcement learning.
The key technical contributions are: 
1) a method to automatically identify a possible hazardous event resulting from the possible actions of another agent and to compute a corresponding disturbance sequence representing the deviation from the nominal predicted behavior,
2) an algorithm to generate a scenario tree with relevant disturbance sequences such that the resulting scenario tree grows linearly with both the prediction horizon and with the number of agents, and
3) an algorithm to prune the resulting scenario tree to obtain a fixed problem size with respect to the number of agents.

\section{Principle of Planner} \label{sec:principle_of_planner}

The proposed planner is based on MPC, which enables straightforward inclusion of constraints and traffic predictions over a finite-time horizon.
To predict the future motion, a discrete model is considered that includes a bounded disturbance $\omega_{k}\in\mathbb{R}^{n_\omega}$ at each discrete time instant.
The dynamics and constraints of the prediction model are described by the nonlinear discrete-time system
\begin{subequations} \label{eq:discrete_system}
\begin{gather}
x_{k+1} = f(x_k,u_k,\omega_k) \label{eq:state_update} \\
h_k(x_k,u_k,\omega_{0:k}) \leq 0 \label{eq:constraint_function}
\end{gather}
\end{subequations}
where $x_k\in\mathbb{R}^{n_x}$ denotes the state,
$u_k\in\mathbb{R}^{n_u}$ denotes the input, $k=0$ corresponds to the
state at the current time instant, and the notation
$\omega_{0:j}=\{\omega_0,\dots,\omega_j\}$ defines that all future
disturbances up to time instant $j$ influence the constraint $h_j$ at time
instant $j$.  Equations \eqref{eq:discrete_system} describe the
ego-vehicle motion and since the constraint function
\eqref{eq:constraint_function} varies with time it can encode
time-dependent constraints, e.g., moving obstacles.

\begin{figure}
\centering
\includegraphics{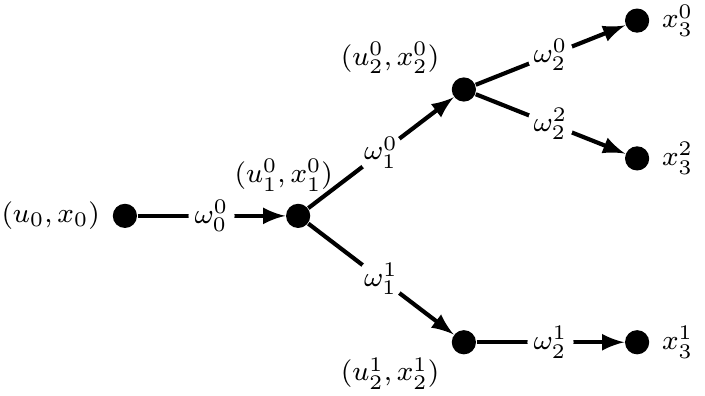}\\
  \caption{\label{fig:nodes}%
    Example of a scenario tree, e.g., arising in traffic with uncertainty.
    The nodes contain the different states $x_k^j$ and control signals $u_k^j$ at different time instants $k$ and realizations $j$.
    A split in the tree occurs when there are more than one disturbance realization $w_k^j$ originating from one node.}
\end{figure}

To handle the disturbances $\omega_{k}$, it would be ideal to have a
control signal $u_k$ for each possible disturbance sequence
$\omega_{0:k}$.
Then, as new measurements become available, the
correct action can be taken with respect to the current disturbance
realization.
For a discrete set of disturbance realizations, each realization
$\omega_k^j$ leads to an alternative state trajectory.
The collection of alternative state trajectories can be
represented by a scenario tree, an example of which is illustrated in
Fig.~\ref{fig:nodes}. 
The point at which a branching occurs in the tree is the earliest
point at which measurements of the actual disturbance realization are
available for feedback.  Specifically, after a disturbance realization
$\omega_k^j$ occurred, the next state $x_{k+1}^j$ and corresponding input $u_{k+1}^j$ can be determined.
The scenario tree is a representation of this presence of feedback in the prediction, which as in feedback MPC \cite{704989} is encoded in the optimal
control problem (OCP) described later in
Sec.~\ref{sec:motion_planning}.

However, there are an infinite number of disturbance realizations and
even if only the extreme values of $\omega_k$ are considered, the
scenario tree grows exponentially with the prediction horizon $N$.
Thus, not all possible disturbance sequences can be covered but there
is still a benefit in considering some key sequences.  This requires
heavy pruning of the scenario tree with a careful selection of
disturbance-sequence realizations and this is described in
Sec.~\ref{sec:disturbance_sequences}.  In short, the determination of
disturbance sequences to include takes advantage of the
re-planning property of MPC by using the planned trajectory at the
previous sample instant.  The strategy is based on finding
disturbance-sequence realizations that make the previously planned
trajectory corresponding to the nominal prediction
($\omega_k=0,\,\forall\,k$) infeasible.  The selection is further
narrowed down to at most a single disturbance-sequence realization per
agent, by selecting the realization based on an adversarial
objective function.

\section{Vehicle and Obstacle Modeling}
In this section, the vehicle equations-of-motion are described and the approach to formulate obstacle constraints is introduced.

\subsection{Kinematic Vehicle Model}
\begin{figure}
\centering
\includegraphics{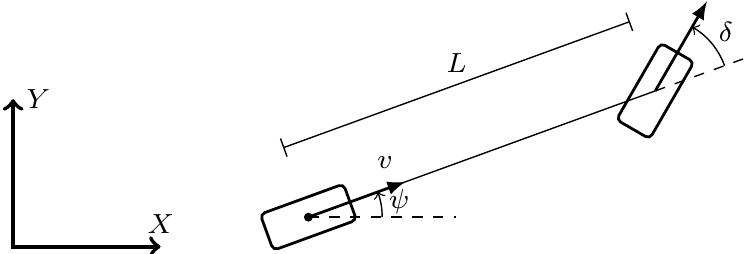}\\
\caption{\label{fig:kinematic_model}%
Kinematic vehicle model.}
\end{figure}

The focus of the motion planning performed in the scenarios considered in this paper is on path and speed planning.
For this purpose a kinematic vehicle model is used, which fulfills constraints imposed by the steering limit.
Considering the vehicle to be operating under normal driving conditions with limited accelerations,
the kinematic model is assumed to sufficiently represent the relevant aspects of the vehicle motion.
To allow for more aggressive maneuvers, a more detailed vehicle model capturing, e.g., the tire--road interaction could be used in the same framework.
The equations of motion for the kinematic vehicle model are \cite{10.5555/1213331}
\begin{subequations} \label{eq:vehicle_dynamics}
\begin{align}
\dot X &= v\cos\psi,&
\dot Y &= v\sin\psi,&
\dot \psi &= v\frac{\tan \delta}{L}\\
\dot v &= u_{a},&
\dot \delta &= u_{\dot\delta}
\end{align}
\end{subequations}
where
$(X,Y)$ is the position of the rear axle in an earth-fixed coordinate frame,
$\psi$ is the vehicle orientation in the same frame,
$v$ is the velocity,
$\delta$ is the steering angle,
$L$ is the wheelbase,
$u_{a}$ is the acceleration input,
and $u_{\dot\delta}$ is the steering-rate input.
The model is illustrated in Fig.~\ref{fig:kinematic_model} as the kinematic single-track model, but it can also represent a traditional vehicle with Ackermann steering geometry.
To obtain discrete-time dynamics of the form \eqref{eq:state_update}, the equations of motion \eqref{eq:vehicle_dynamics} are discretized using the Runge-Kutta fourth-order method RK4 \cite{ascher}.

\subsection{Simplified Vehicle Model}

For the purpose of fast lattice planning to compute disturbance sequences (see Sec.~\ref{sec:computing_disturbance_sequences}) and an initial guess for numerical optimization (see Sec.~\ref{sec:initialization}), a simpler vehicle model is also used.
This model is obtained by setting $\delta=\psi=0$, where the lateral movement is instead determined by the lateral movement input $u_y = \dot Y/\dot X$.
Written with $X$ as the free variable, $Y$, $v^2$, and the time $t$ as states, and $u_a$ and $u_y$ as constant inputs for each change $\Delta X$ of $X$, the model can be written in discrete time as:
\begin{subequations} \label{eq:simple_vehicle_model}
\begin{align}
X_{i+1} &= X_i + \Delta X \\
Y_{i+1} &= Y_i + u_y \Delta X \\
v_{i+1}^2 &= v_i^2 + 2 u_a \Delta X \\
t_{i+1} &= t_i + \begin{cases}
-\dfrac{v_i}{u_a} + \sqrt{\dfrac{v_i^2}{u_a^2}+\dfrac{2\Delta X}{u_a}}, & u_a > 0 \\
\dfrac{\Delta X}{v_i}, & u_a = 0. \\
-\dfrac{v_i}{u_a} - \sqrt{\dfrac{v_i^2}{u_a^2}+\dfrac{2\Delta X}{u_a}}, & u_a < 0
\end{cases}
\end{align}
\end{subequations}

\subsection{Obstacle Constraints}

\begin{figure}
\centering
\includegraphics{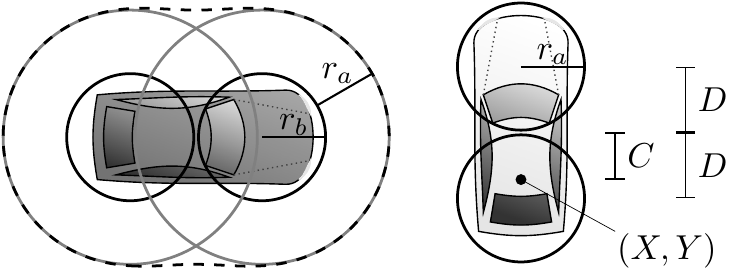}\\
\caption{\label{fig:obstacle}%
Obstacle constraints with associated radii and distances involved in the modeling.}
\end{figure}

One circle is not suitable to represent the vehicle shape and
orientation. Therefore, the obstacle constraints for the vehicles are
formulated using a kernel representation based on two circles as seen
in Fig.~\ref{fig:obstacle}.  
This will ensure a smooth representation
of the rectangular-shaped vehicle.  The distance from the reference
position $(X,Y)$ at the rear axle to the center of the circles are determined by the distance $C$ to
the vehicle body center, and the distance $D$ from the center of the vehicle body to the center of each circle.  The positions of the center points $(X_1, Y_1)$
and $(X_2, Y_2)$ of the respective circle are determined by the
relations
\begin{subequations}
\begin{align}
X_1 &= X + (C+D)\cos(\psi) \\
Y_1 &= Y + (C+D)\sin(\psi) \\
X_2 &= X + (C-D)\cos(\psi) \\
Y_2 &= Y + (C-D)\sin(\psi).
\end{align}
\end{subequations}
As illustrated by the two larger gray circles in Fig.~\ref{fig:obstacle}, constraints between the center points $a$ and $b$ of two circles with the radii $r_a$ and $r_b$, respectively,
can be viewed as a larger circle constraining the point $b$ a distance $r_a+r_b$ away from $a$ and be formulated as
\begin{equation}
\norm{a-b}_2^2 - (r_a + r_b - s_k)^2 \geq 0
\end{equation}
where $s_k$ is introduced as a slack variable.
For a more vehicle-shaped constraint than the two circles give, the constraints for both points on one vehicle are combined using the squared exponential kernel function with the length scale $l$.
With points $a_i, \, i = 1,\ldots,N_i$, on one vehicle to compare to each point $b_j, \, j=1,\ldots,N_j$, on another  vehicle, the constraints are formulated as
\begin{subequations} \label{eq:kernel_collision_constraint}
\begin{gather}
\sum_{i=1}^{N_i} k(a_i,b_j) \leq 1, \, \forall \, j \label{eq:obstacle_constraint_first}\\
	k(a,b) = \exp\left(-\frac{\norm{a-b}_2^2 - (r_a + r_b - s_k)^2}{2l^2}\right).
\end{gather}
\end{subequations}
The resulting constraint for each circle center point $a_1$ and $a_2$ on the black vehicle in Fig.~\ref{fig:obstacle} is illustrated as a dashed black line overlaying the gray vehicle.
For each target vehicle, the constraints are combined and compared to the center point of the circles overlaying the ego vehicle.
The length scale $l$ is selected such that the constraint boundary is at least the distance $r_a+r_b$ from the vehicle centerline:
\begin{equation}
l = \frac{D}{\sqrt{2 \ln{2}}}.
\end{equation}
The lane constraints are enforced as
\begin{subequations} \label{eq:lane_constraint}
\begin{align}
Y_{j,k} - Y_\mathrm{min}(X_{j,k}) + s_k &\geq r_j \\
Y_\mathrm{max}(X_{j,k}) - Y_{j,k} + s_k &\geq r_j, \, \forall\, j, k \label{eq:obstacle_constraint_last}
\end{align}
\end{subequations}
where $Y_\mathrm{min}$ and $Y_\mathrm{max}$ are functions of $X_{j,k}$ describing the lane limit along the $Y$-direction.

\section{Determining Disturbance Sequences} \label{sec:disturbance_sequences}

Even if it was computationally feasible to consider all possible
disturbance sequences, it may not be desirable as the solution will be
overly conservative.  One way to deal with this is to introduce rules
such as existing traffic laws, but other agents on the road would not
necessarily follow the anticipated rules and traffic laws do not
specify all interactions between traffic participants.  There could be
many traffic situations where there is a need for negotiation or to
predict the risk of performing a maneuver to drive safely.  A
risk-averse strategy is here formulated by considering how other
vehicles could disturb the ego vehicle if it drives according to its
nominal behavior.

\subsection{Building the Scenario Tree} \label{sec:building_tree}

To enable long prediction horizons, it is essential to only consider a
carefully chosen subset of the possible disturbance realizations that
influence the ego-vehicle solution.  Even with feedback of the ego
vehicle present in the predictions, an adversarial disturbance model
can result in overly conservative solutions or even infeasible
problems if worst-case behavior of surrounding traffic is considered.
To reduce the conservatism compared to the worst-case adversarial
disturbance model, it is assumed that other agents are not of ill
intent, i.e., they will not negatively feedback on the motion of the
ego vehicle to cause a collision but can still deviate significantly
from their predicted behavior.  This is formalized as an open-loop
adversarial model of the disturbances.
\begin{definition}[Open-loop adversarial]
  A disturbance-sequence realization $\omega_{0:N}$ is open-loop
  adversarial with respect to an initial state $\hat x_0$ and an input
  sequence $\hat u_{0:N}$, if it is bounded 
  and $\exists\, j\in \{0,\dots,N\}$ such that for \eqref{eq:discrete_system},
  $h_j(x_j,u_j,\omega_{0:j})>0$, under the motion equations
  $x_{k+1} = f(x_k,u_k,\omega_k)\,\forall\, k\in \{0,\dots,N\},\, x_0=\hat x_0,\text{ and }
  u_{0:N}=\hat u_{0:N}$.
\end{definition}

Out of the in total $n_\omega$ disturbances, each agent $a$ has $n_a$
disturbances
$\omega^a\in\mathbb{R}^{n_a}\subseteq \mathbb{R}^{n_\omega}$
explicitly acting on it, e.g., deviations from the predicted actions
of the agent such as unexpected acceleration or steering.  Now, the
objective is to formulate an algorithm that for each agent $a$
collects a single open-loop adversarial sequence $\omega_{0:N}$ that
explicitly acts on agent $a$.  With $N_a$ agents, the worst case
computationally is that all $N_a$ disturbance sequences obtained with
such an algorithm start from the root node and the resulting scenario
tree (see Fig.~\ref{fig:nodes}) contains $N + N_a (N-1)$ nodes.  Note
that this tree grows linearly with both the prediction horizon and
with the number of agents and therefore the approach does not suffer
from exponential complexity growth with increasing problem size.

The input to the algorithm is an estimation of the current state
$\hat x_0$, the set of agents $\mathcal{A}$, and a trajectory tree
$\mathcal{T}$ containing predicted inputs $\hat u$ obtained from the
previous planning iteration by the MPC planner (see
Sec.~\ref{sec:motion_planning}).  The output is a disturbance tree
$\mathcal{W}$, describing how the disturbance sequences associated with
each agent branch off the nominal disturbance sequence $\omega_{0:N}=0$.
The algorithm proceeds as follows: 1) from the previously planned
trajectory tree $\mathcal{T}$, the nominal input trajectory
$\hat u_{0:N}$ corresponding to no disturbance ($\omega_{0:N}=0$) is
extracted ($\mathrm{extract\_nominal}(\mathcal{T})$); 2) for each agent $a$ in $\mathcal{A}$; 3) search
for an open-loop adversarial disturbance sequence
$\omega_{k:N}, k\in \{0,\dots N\}$ that explicitly acts only on agent $a$, 
i.e., only the subspace of disturbances
$\omega^a\in\mathbb{R}^{n_a}\subseteq \mathbb{R}^{n_\omega}$ in $\omega\in\mathbb{R}^{n_\omega}$ are nonzero.
($\mathrm{get\_open\_loop\_adversarial}(a, \hat x_0, \hat u_{0:N})$);
and 4) if such a sequence exists; then 5) add it to the disturbance tree
$\mathcal{W}$.  The algorithm is summarized by:
\begin{algorithmic}[1]
	\HEADER Generate disturbance tree $\mathcal{W}$
	\algrule
	\INPUT Initial state $\hat x_0$, Agents $\mathcal{A}$, Trajectory tree $\mathcal{T}$
	\OUTPUT Disturbance tree $\mathcal{W}$
	\State $\hat u_{0:N} \leftarrow \mathrm{extract\_nominal}(\mathcal{T})$
	\ForEach{$a \in \mathcal{A}$}
		\State $\omega_{k:N} \leftarrow \mathrm{get\_open\_loop\_adversarial}(a, \hat x_0, \hat u_{0:N})$
		\If{$\omega_{k:N} \ne \emptyset$}
			\State Add $\omega_{k:N}$ to $\mathcal{W}$
		\EndIf
	\EndFor
	\algrule
\end{algorithmic}
The complete scenario tree with states, inputs, and disturbances is formed by
associating the resulting disturbance tree $\mathcal{W}$ and the nominal disturbance-free sequence $\omega_{0:N}=0$ with the corresponding states and inputs as illustrated in Fig.~\ref{fig:nodes}.

\subsection{Adversarial Objective Function} \label{sec:adversarial_objective_function}

As stated, there are possibly infinitely many open-loop adversarial
disturbance-sequence realizations, so a criterion is needed to
determine which to consider.  One important variable is the time,
$t_\mathrm{inf}$, indicating when the planned trajectory becomes
infeasible. Before formal definitions, let $\mathcal{D}$ denote the
set of all open-loop adversarial disturbance sequences and
$t:\mathbb{N}\rightarrow\mathbb{R}$ be a map from a discrete-time
index to the corresponding time.
\begin{definition}[$t_\mathrm{inf}$]
  For a disturbance $\omega_{0:N}\in\mathcal{D}$, then
  $t_\mathrm{inf}(\omega_{0:N})=t(k)$, where $k\in\{0,\dots,N\}$ is the smallest $k$
  such that $\omega_{0:k}\in\mathcal{D}$.
\end{definition}
It is reasonable to assume that a disturbance sequence that makes
the planned trajectory infeasible earlier is more difficult to
manage by re-planning.  Similarly, a disturbance sequence that
deviates from the nominal behavior later results in less time
available to compensate for it. Therefore, another important variable is the
starting time, $t_\mathrm{dist}$, of the disturbance sequence.
\begin{definition}[$t_\mathrm{dist}$]
  For a disturbance $\omega_{0:N}\in\mathcal{D}$, then
  $t_\mathrm{dist}(\omega_{0:N})=t(k)$, where $k\in\{0,\dots,N\}$ is the smallest $k$ such that $\norm{\omega_k} \ne 0$.
\end{definition}
Further, the magnitude of the resulting constraint violation and
possibly the increase in objective cost could be considered. Thus,
there is a trade-off between making the planned trajectory becoming
infeasible early against a disturbance sequence starting
later. Therefore, a trade-off parameter $\eta\in[0, 1)$ is introduced
in a minimization problem with the adversarial objective function
\begin{equation} \label{eq:oppositional_objective}
\minimize_{\omega_{0:N}}
t_\mathrm{inf}(\omega_{0:N}) - \eta\,t_\mathrm{dist}(\omega_{0:N}), \, \omega_{0:N} \in \mathcal{D}.
\end{equation}
At time $t_\mathrm{inf}$, the future disturbance sequence is determined by a minimum-time solution that intentionally increases the resulting constraint violation (see Sec.~\ref{sec:computing_disturbance_sequences}).

By design, the disturbance tree $\mathcal{W}$ obtained using the algorithm in Sec.~\ref{sec:building_tree} grows linearly with the number of agents.
To obtain a fixed tree size with respect to the number of agents, the tree $\mathcal{W}$ can be pruned to only include the $n$ most critical disturbance sequences.
The adversarial objective function \eqref{eq:oppositional_objective} can be used for this purpose,
where the pruned tree is obtained by the following algorithm:
\begin{algorithmic}[1]
  \HEADER Prune disturbance tree $\mathcal{W}$
  \algrule
  \INPUT Disturbance tree $\mathcal{W}_\mathrm{in}$, Number of disturbance sequences $n$
  \OUTPUT Disturbance tree $\mathcal{W}_\mathrm{out}$
  \State Sort each $\omega_{k:N}$ in $\mathcal{W}_\mathrm{in}$ by \eqref{eq:oppositional_objective}
  \State $n \leftarrow \min(n,\mathrm{number\_of\_sequences\_in}(\mathcal{W}_\mathrm{in}))$
  \State Add the $n$ first $\omega_{k:N}$ in $\mathcal{W}_\mathrm{in}$ to $\mathcal{W}_\mathrm{out}$
  \algrule
\end{algorithmic}

\subsection{Computing Disturbance Sequences} \label{sec:computing_disturbance_sequences}

To compute the disturbance sequence that minimizes
\eqref{eq:oppositional_objective} for a specific target vehicle, a fast search on a discretized
state lattice is performed using Dijkstra's algorithm
\cite{10.5555/1213331}.  Collisions with vehicles other than the ego
vehicle are ignored, because this decouples the motion of each
vehicle.  Further, this choice allows to severely reduce the search
space by not including the time $t$ as a state on the lattice.

Using the simplified vehicle model \eqref{eq:simple_vehicle_model}, the lattice is formed on the variables $X$, $v^2$, and $t_\mathrm{dist}$ where the
possible actions are $u_a=\pm a_\mathrm{dist}$, or to keep on the nominal
predicted trajectory.
The lateral position $Y_\mathrm{TV}$ of the target vehicle is determined in a feedback fashion such that the target vehicle approaches the lateral position $Y_\mathrm{ego}$ of the ego vehicle by a fraction $k_y$ of its longitudinal movement
\begin{equation}
u_y = k_y \sign(Y_\mathrm{ego}-Y_\mathrm{TV}).
\end{equation}
Infeasibility of the ego-vehicle trajectory is
determined by the collision constraint
\eqref{eq:kernel_collision_constraint} being unfulfilled without a
slack $s_k$.
Only open-loop adversarial disturbance-sequence realizations within the prediction horizon of the MPC are considered.
When an adversarial disturbance-sequence realization is found, 
a minimum-time based control law 
is used to intentionally keep the constraint violation large for the remainder of the prediction horizon.
The control law applies the acceleration on the target vehicle that does not reduce the velocity if it is below $v_\mathrm{min}$, and will bring the target vehicle to the ego vehicle in minimum time if the ego vehicle keeps a constant velocity:
\begin{subequations}
\begin{align}
  x &= -\dfrac{|v_\mathrm{TV}-v_\mathrm{ego}|(v_\mathrm{TV}-v_\mathrm{ego})}{2 a_\mathrm{dist}} \\
  a_\mathrm{TV} &= \begin{cases}
    a_\mathrm{dist}, & X_\mathrm{TV}-X_\mathrm{ego} < x \text{ and } \\
    & v_\mathrm{TV} + a_\mathrm{dist} \Delta t^2/2 < v_\mathrm{min} \\
    -a_\mathrm{dist}, & \text{otherwise}
  \end{cases}
\end{align}
\end{subequations}
where $v_\mathrm{TV}-v_\mathrm{ego}$ is the difference in speed between the target vehicle and the ego vehicle.

\subsection{Importance Weights} \label{sec:importance_weights}

When formulating the motion planner objective function, see
\eqref{eq:scenario_ocp} in Sec.~\ref{sec:motion_planning}, it is of
interest to assign importance weights $\beta$ to the different branches in the
scenario tree. 
  If the weights were to represent the true
  statistical distribution, then the objective function
  \eqref{eq:scenario_ocp} is equivalent to minimizing the
  expected cost.
Without knowing the true distribution,
the challenge is how the weights should be
assigned. The selection of weights are based on the following three
principles:
\begin{myrule} \label{assump:weights1}
The weights of all nodes corresponding to the same time instant sum up to 1.
\end{myrule}
\begin{myrule} \label{assump:weights2}
Adding a branch retains the relative weights between existing branches.
\end{myrule}
\begin{myrule} \label{assump:weights3}
The weights of outgoing branches from a node sum up to the incoming weight to the node.
\end{myrule}
Principle~\ref{assump:weights2} means that the more branches that are
included in the problem, the less likely each individual branch is.
Principle~\ref{assump:weights3} means that the weights remain
consistent such that for any sub-tree of the scenario tree, the sum of
the weights at each time instant is equal, which is what would be
expected if interpreting the weights as probabilities.  To fulfill
Principles~\ref{assump:weights1}--\ref{assump:weights3} and to enable some branches to be prioritized, the odds parameter $\gamma$ is introduced to compute the importance weights of a scenario tree.
The weight $\beta_{j}$ for each child branch $j$ from node $i$ is computed as
\begin{equation}
  \beta_j=\beta_i\frac{\gamma_j}{\sum_k \gamma_k}
\end{equation}
where $k$ is the number of outgoing branches from node $i$.

\begin{figure}
\centering
\includegraphics{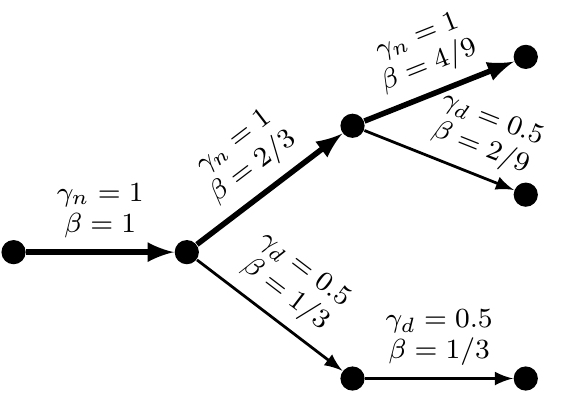}\\
\caption{\label{fig:weights}%
Illustration of weights for each node in an example scenario tree.
Each edge in the tree is parameterized with the odds parameter $\gamma$,
resulting in the different importance weights $\beta$.
}
\end{figure}

An example of a tree with assigned weights is illustrated in Fig.~\ref{fig:weights}.
To parameterize the tree, the nominal branch is given the odds parameter
$\gamma_n = 1$ and the disturbance branches are given the odds parameter $\gamma_d$.
The odds parameter $\gamma_d$ should be considered a tuning parameter of how
much, apart from being feasible, should the alternative branches
influence the resulting optimization result.  A large $\gamma_d$ can
make the problem more robust to disturbance-sequence realizations that
are different, but similar, to those included in the scenario tree,
while a smaller $\gamma_d$ improves the performance under nominal
conditions where the other agents behave as expected.

\section{Motion Planning and Control} \label{sec:motion_planning}

With the adversarial disturbance-sequence realization and the
associated scenario tree
computed according to Sec.~\ref{sec:disturbance_sequences}, the motion
planner and controller is defined as an MPC problem, resulting in the
Adversarial Disturbance-Sequence Branching MPC (ADSB-MPC) strategy.

In the nominal case, where it is assumed that there are no
disturbances, the MPC problem is formulated as
\begin{subequations} \label{eq:nmpc}
  \begin{alignat}{2}
    \minimize_{x,u,s} \enspace 
    & \mathrlap{\sum_{k=1}^{N+1} \bigg( \begin{bmatrix}\Delta x_{k}\\u_{k-1}\end{bmatrix}^\mathrm{T}\! Q \begin{bmatrix}\Delta x_{k}\\ u_{k-1}\end{bmatrix}
      + b\, s_{k}\bigg)} \label{eq:nmpc1}\\
    \mathrm{subject~to} \enspace 
    & x_0 = \hat x_0 \label{eq:nmpc2}\\
    & x_{k} = f(x_{k-1},u_{k-1}), \; &&k=1,\dots,N+1 \label{eq:nmpc3}\\
    & g(x_{k}, o_k, s_k) \leq 0, \; &&k=1,\dots,N+1 \label{eq:nmpc4}\\
    & x_\mathrm{min} \leq x_{k} \leq x_\mathrm{max}, \; &&k=1,\dots,N+1 \label{eq:nmpc5}\\
    & u_\mathrm{min} \leq u_k \leq u_\mathrm{max}, \; &&k=0,\dots,N \label{eq:nmpc6}\\
    & s_k \geq 0, \; &&k=1,\dots,N+1 \label{eq:nmpc7}
  \end{alignat}
\end{subequations}
where the objective function \eqref{eq:nmpc1} is chosen as a quadratic stage cost where $\Delta x_{k}=x_k-x_{\mathrm{ref}}$ is the state error relative to a low-complexity reference $x_{\mathrm{ref}}$ specifying the desired speed and lane, the matrix $Q$ is positive semi-definite and $b$ is a scalar,
the constraint \eqref{eq:nmpc2} enforces that the prediction starts at the estimated current state $\hat x_0$,
the constraints \eqref{eq:nmpc3} enforce the disturbance-free dynamics \eqref{eq:vehicle_dynamics},
the constraints \eqref{eq:nmpc4} enforce the disturbance-free obstacle constraints \eqref{eq:kernel_collision_constraint}--\eqref{eq:lane_constraint} with predicted obstacle positions $o_k$,
the constraints \eqref{eq:nmpc5} enforce the state limits such as maximum speed,
the constraints \eqref{eq:nmpc6} enforce the input limits such as maximum acceleration,
and the constraints \eqref{eq:nmpc7} enforce that the slack variables $s_k$ are non-negative.

The corresponding branching optimal control problem that considers the adversarial disturbance-sequence realizations differs only in a few aspects to \eqref{eq:nmpc}.
To describe the scenario tree $\mathcal{T}$, illustrated in Fig.~\ref{fig:nodes}, the functions $p_x:\mathbb{R}^{n_x}\rightarrow \mathbb{R}^{n_x}$, and $p_u:\mathbb{R}^{n_x}\rightarrow\mathbb{R}^{n_u}$ are introduced.
The function $p_x(x_k)$ returns the state of the parent node to state $x_k$ and the function $p_u(x_k)$ returns the corresponding input that brings the system from $p_x(x_k)$ to $x_k$. 
For example, using the notation in Fig.~\ref{fig:nodes}, this gives $p_x(x_{3}^2)=x_{2}^0$ and $p_u(x_{3}^2)=u_{2}^0$.
The total number of state-trajectory elements $1+N_x$ is equal to $1+\sum_b N_b$, where $N_b$ is the length of each branch from its parent and the first term accounts for $x_0$. The total number of input-trajectory elements $1+N_u$ is equal to $1+\sum_b (N_b-1)$ (the terminal state elements do not have corresponding inputs).
Flattening the index notation such that $x_k,\, k=0,\dots,N_x$, are all the states in the scenario tree and $u_k,\, k=0,\dots,N_u$, are all the inputs, the ADSB-MPC optimal control problem is formulated as
\begin{subequations}\label{eq:scenario_ocp}%
\begin{alignat}{2}
\minimize_{x,u,s} \enspace
& \mathrlap{\sum_{k=1}^{N_x} \beta_k \bigg( \begin{bmatrix}\Delta x_{k}\\p_u(x_{k})\end{bmatrix}^\mathrm{T}\! Q \begin{bmatrix}\Delta x_{k}\\p_u(x_{k})\end{bmatrix}
+ b\, s_{k}\bigg)} \label{eq:smpc1} \\
\mathrm{subject~to} \enspace 
& x_0 = \hat x_0 \label{eq:smpc2} \\
& x_{k} = f(p_x(x_k),p_u(x_k)), \; &&k=1,\dots,N_x \label{eq:smpc3} \\
& g(x_k,o_k,s_{k}) \leq 0, \; &&k=1,\dots,N_x \label{eq:smpc4} \\
& x_\mathrm{min} \leq x_k \leq x_\mathrm{max}, \; &&k=1,\dots,N_x \label{eq:smpc5} \\
& u_\mathrm{min} \leq u_k \leq u_\mathrm{max}, \; &&k=0,\dots,N_u \label{eq:smpc6} \\
& s_{k} \geq 0, \; &&k=1,\dots,N_x
\end{alignat}
\end{subequations}
The importance weights $\beta_k$ enter the objective function to prioritize minimizing the stage cost of branches with higher importance weight (see Sec.~\ref{sec:importance_weights}).
The scenario tree is encoded in the formulation by the dynamic constraints \eqref{eq:smpc3}.
Only disturbances affecting the obstacle positions $o_k$ are considered and they are encoded in \eqref{eq:smpc4} by different predicted positions $o_k$ for each $k=1,\dots,N_x$.
In general, disturbances affecting the dynamics \eqref{eq:state_update} can be encoded by different dynamic equations $f_k$ for each $k=1,\dots,N_x$, and 
disturbances affecting the constraints \eqref{eq:constraint_function} can be encoded by different constraint equations $h_k$ for each $k=1,\dots,N_x$.

The procedure for ADSB-MPC can be summarized as follows.
At each sample, the input trajectory $\hat u_{0:N}$ corresponding to the disturbance-free branch is extracted from the solution to \eqref{eq:scenario_ocp} at the previous sample and
a new scenario tree with associated disturbance sequences is obtained according to the algorithm described in Sec.~\ref{sec:building_tree} with importance weights computed according to Sec.~\ref{sec:importance_weights}.
Each disturbance sequence building up the scenario tree is computed using the adversarial objective function described in Sec.~\ref{sec:adversarial_objective_function} and the procedure in Sec.~\ref{sec:computing_disturbance_sequences}.
The optimization problem \eqref{eq:scenario_ocp} is then re-solved with this new scenario tree and the control input obtained at the root of the resulting trajectory tree is applied to the system, and then this process is repeated at the next sample.

\subsection{Initialization} \label{sec:initialization}

Since the problems \eqref{eq:nmpc}--\eqref{eq:scenario_ocp} are non-convex, the
initialization of the variables in the optimal control problem is of high importance
when considering numerical solution approaches.  Solving
\eqref{eq:nmpc} or \eqref{eq:scenario_ocp} using a local optimization
method means that the solver is not guaranteed to converge to a global
optimum.
To initialize the problem, it is therefore first approximated and
solved using a global optimization method, in this case a search on a
discretized state lattice using A* \cite{10.5555/1213331}.  Similar to when finding adversarial
disturbance sequences in Sec.~\ref{sec:disturbance_sequences}, the
simplified vehicle dynamics \eqref{eq:simple_vehicle_model} are discretized on a
lattice.  Because the solution to \eqref{eq:nmpc} and
\eqref{eq:scenario_ocp} are dependent on moving obstacles and the
lateral vehicle motion, the lattice is formed on the states $X$, $Y$,
$v^2$, and $t$. While the time $t$ does not conform to a lattice using
\eqref{eq:simple_vehicle_model}, the search graph is reduced by forming cells
with discretization $\Delta t$, allowing at most one
vertex per cell for each unique $X$, $Y$, and $v^2$.

In total there are 5
actions available: keep constant velocity,
accelerate with $a_\mathrm{max}$, brake with $-a_\mathrm{max}$, increase the lateral position, and decrease the lateral position.  While the discretization of the input space is coarse to make the search real-time feasible, 
the trajectories are refined by
solving \eqref{eq:nmpc} and \eqref{eq:scenario_ocp} using direct
optimal control, where the variables are initialized with the result from the lattice search.  Compared to dynamic-programming
approaches such as lattice search, direct optimal control scales well
with respect to the prediction horizon and the discretization
resolution of states and inputs.

To initialize the branching trajectory in \eqref{eq:scenario_ocp}, each disturbance sequence included in the problem is solved for individually using the lattice search.
This means that in practice it is assumed that the global minimum does not significantly change when simultaneously
optimizing for all branches, e.g., the optimal solution for a branch does not move from merging in front of a specific target vehicle to merging behind it.

\section{Results} \label{sec:results}

Prototype implementations of the Nominal MPC \eqref{eq:nmpc} and the ADSB-MPC \eqref{eq:scenario_ocp} strategies were made in Python and C++, using CasADi \cite{casadi} to formulate the optimal control problems, which were subsequently solved numerically by the interior point optimizer Ipopt \cite{ipopt} together with the linear systems solver MA57 \cite{hsl}.
The ADSB-MPC is first evaluated in two hand-crafted merging scenarios and then in a
large number of randomized highway-driving scenarios.
A video illustrating the different simulations is available at \url{https://youtu.be/etxvjGoX7RA}.

Table~\ref{tab:parameters} summarizes the parameters used in the
evaluations, where
$w_\mathrm{lane}$, $L$, $r$, $D$, $C$, $v_\mathrm{ref}$, $v_\mathrm{max}$, $v_\mathrm{min}$, $\psi_\mathrm{max}$, $\delta_\mathrm{max}$ and $a_\mathrm{max}$ are vehicle and scenario parameters.
The reference $x_{\mathrm{ref}}$ is to drive straight in the center of the target lane with the speed $v_\mathrm{ref}$.
The values of $Q$ and $b$ in \eqref{eq:nmpc} and \eqref{eq:scenario_ocp} are chosen such that the states $x_k=[X_k,Y_k,\psi_k,v_k,\delta_k]^\mathrm{T}$, the inputs $u_k=[u_{k,\dot\delta},u_{k,a}]^\mathrm{T}$, and the slack $s_k$ give the same contribution to the objective function for a deviation equal to their specified cost normalization (as per Table~\ref{tab:parameters}) :
\begin{align}
Q &= \diag(0, q_Y^{-2}, q_\psi^{-2}, q_v^{-2}, q_\delta^{-2}, q_{\dot \delta}^{-2}, q_a^{-2}), &
b&=q_s^{-1}.
\end{align}
The cost-normalization values are first selected based on worst case
(similar to Bryson's rule \cite{franklin2009feedback}) and then
$q_\delta$, $q_{\dot\delta}$ were tuned to obtain smooth steering
and $q_v$, $q_a$ were tuned to obtain reasonable accelerations.  In
the highway scenario, the steering-rate cost $q_{\dot\delta}$ was set to
$\infty$, since in that simulation the steering angle
was directly used as the input.  The parameters $\gamma_d$, $\eta$,
$a_\mathrm{dist}$, and $k_y$ are specific to the ADSB-MPC.  The values
$0.25$, $0.5$, and $0.75$ were tested for $\gamma_d$ in the merging
scenario, which gave similar results but different prioritization for
the branches.  The adversarial objective weight $\eta$ is selected as
a small value to prioritize smaller $t_\mathrm{inf}$, but large enough
such that for disturbance sequences with similarly sized $t_\mathrm{inf}$
the one with the largest $t_\mathrm{dist}$ is selected. In
simulations, it has proven that the results are not very sensitive to the
particular choice of $\eta$. The acceleration disturbance $a_\mathrm{dist}$ was
lowered for the highway scenario to promote overtaking also in
scenarios with dense traffic.  The lateral disturbance parameter $k_y$
was set to zero in the merging scenario, as the other agents were
restricted to a single lane.

\begin{table}[!t]
\renewcommand{\arraystretch}{1.3}
\setlength{\tabcolsep}{5pt}
\caption{Parameters}
\label{tab:parameters}
\centering
\begin{tabular}{lcccc}
\hline
\multirow{2}{*}{\bfseries Description} & \multirow{2}{*}{\bfseries Notation} & \multicolumn{2}{c}{\bfseries\underline{Scenario}} & \multirow{2}{*}{\bfseries Unit} \\
& & \bfseries Merge & \bfseries Highway & \\
\hline
Lane width & $w_\mathrm{lane}$ & 3 & 4 & m \\
Wheelbase & $L$ & 2 & 5 & m \\
Collision circle radius & $r$ & 1.35 & 1.4 & m \\
Collision circle separation & $D$ & 1.4 & 1.4 & m \\
Vehicle body center & $C$ & 1 & 2.5 & m \\
Maximum speed & $v_\mathrm{max}$ & 30 & 30 & m/s \\
Minimum speed & $v_\mathrm{min}$ & 0 & 0 & m/s \\
Maximum orientation & $\psi_\mathrm{max}$ & $\pi/2$ & $\pi/2$ & rad \\
Maximum steering angle & $\delta_\mathrm{max}$ & $\pi/6$ & $\pi/4$ & rad \\
Maximum acceleration & $a_\mathrm{max}$ & 3 & 5 & m/s$^2$ \\
Reference speed & $v_\mathrm{ref}$ & 25 & 30 & m/s \\
Lateral position cost & $q_{Y}$ & 3 & 12 & m \\
Orientation cost & $q_{\psi}$ & $\pi/2$ & $\pi/2$ & rad \\
Speed cost & $q_{v}$ & 10 & 20 & m/s \\
Steering angle cost & $q_{\delta}$ & 0.1 & 0.1 & rad \\
Steering rate cost & $q_{\dot\delta}$ & 0.1 & $\infty$ & rad/s \\
Acceleration cost & $q_{a}$ & 6 & 10 & m/s$^2$ \\
Slack cost & $q_{s}$ & 0.001 & 0.001 & m \\
MPC prediction horizon & $N$ & 25 & 15 & - \\
MPC sample time & - & 0.2 & 0.2 & s \\
Lattice position discretization & $\Delta X$ & 5 & 5 & m \\
Lattice time discretization & $\Delta t$ & 0.01 & 0.01 & s \\
Branch odds & $\gamma_d$ & 0.5 & 0.5 & - \\
Adversarial objective weight & $\eta$ & 0.25 & 0.25 & - \\
Acceleration disturbance  & $a_\mathrm{dist}$ & 3 & 1 & m/s$^2$ \\
Lateral disturbance & $k_y$ & 0 & 0.1 & - \\
\hline
\end{tabular}
\end{table}

\subsection{Merging}

To make the merging scenario feasible, it is in the generation of disturbance sequences assumed that a target vehicle will not actively try to crash into the ego vehicle if the ego vehicle has merged into the same lane as the target vehicle, which is determined by the condition
\begin{equation}
|Y_\mathrm{TV}-Y_\mathrm{ego}| \leq \frac{w_\mathrm{lane}}{2}
\end{equation}
where $Y_\mathrm{TV}$ is the lateral position of the target vehicle, 
$Y_\mathrm{ego}$ is the lateral position of the ego vehicle,
and $w_\mathrm{lane}$ is the lane width.
Two different merging scenarios are examined: a disturbance-free scenario, where the other vehicles in the traffic are driving with constant velocity,
and a disturbance scenario, where the white target vehicle performs a 3\,m/s$^2$ acceleration, followed by a deceleration phase.

\subsubsection{Disturbance-Free Scenario}

\begin{figure}
\centering
\includegraphics{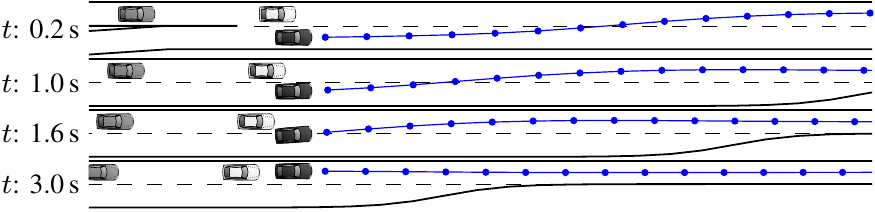}\\
\caption{\label{fig:nominal}%
Disturbance-Free Scenario: Black ego vehicle using the Nominal MPC strategy at different time instants. The future planned positions are illustrated with dots (with a sample period of 0.2\,s).}
\end{figure}

\begin{figure}
\centering
\includegraphics{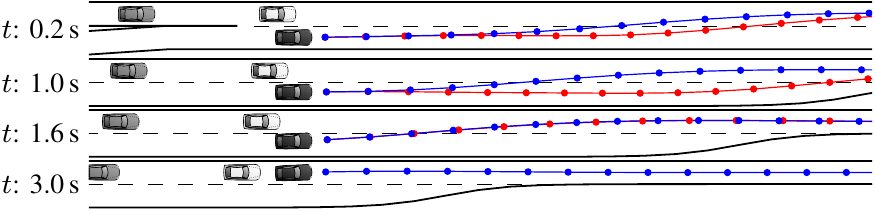}\\
\caption{\label{fig:branching}%
Disturbance-Free Scenario: Black ego vehicle using the ADSB-MPC strategy at different time instants. The future planned positions are illustrated with dots (with a sample period of 0.2\,s).}
\end{figure}

In the disturbance-free merging scenario, the target vehicles behave as the nominal prediction in the MPC
controllers, i.e., drive with a constant velocity.  As seen in
Figs.~\ref{fig:nominal} and \ref{fig:branching}, both the Nominal MPC
and the proposed ADSB-MPC are able to successfully perform a merging
maneuver in this scenario.  At early stages of the scenario, the
ADSB-MPC has an alternative branched trajectory where, if the white
car were to accelerate, the black ego vehicle would merge behind the
white vehicle, which is seen in the alternative lower plans in the top
illustrations in Fig.~\ref{fig:branching}.  Later, once the ego
vehicle has high enough relative velocity and is far enough in front
of the white vehicle, the alternative plan is to simply accelerate
more to avoid collision in case the white vehicle performs an
acceleration.  Throughout the maneuver, there is no disturbance
sequence affecting the gray vehicle that would prevent the black ego
vehicle from merging.

\subsubsection{Disturbance Scenario}

\begin{figure}
\centering
\includegraphics{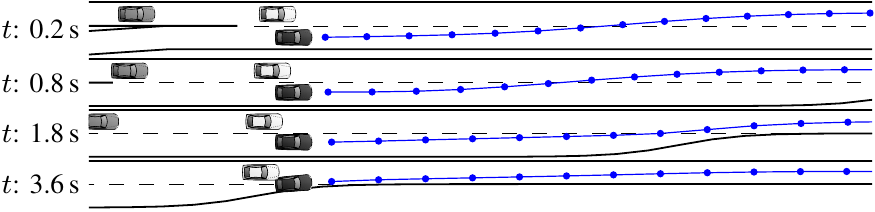}\\
\caption{\label{fig:nominal_dist}%
Disturbance Scenario: Black ego vehicle using the Nominal MPC planner at different time instants. The future planned positions are illustrated with dots (with a sample period of 0.2\,s).}
\end{figure}

\begin{figure}
\centering
\includegraphics{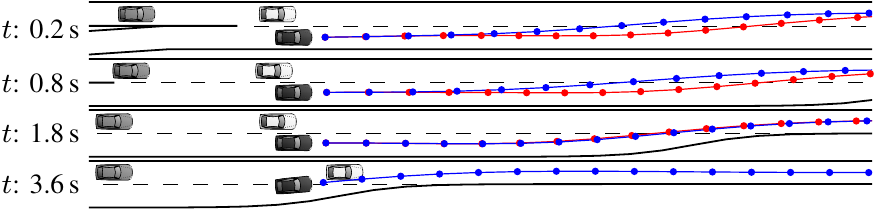}\\
\caption{\label{fig:branching_dist}%
Disturbance Scenario: Black ego vehicle using the ADSB-MPC planner at different time instants. The future planned positions are illustrated with dots (with a sample period of 0.2\,s).}
\end{figure}

\begin{figure}
\centering
\includegraphics{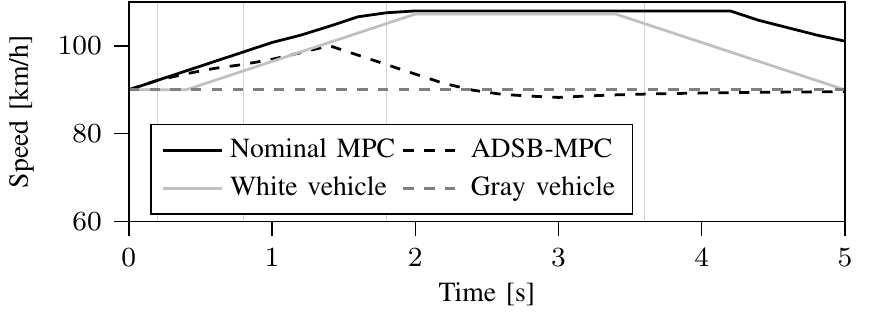}\\
\caption{\label{fig:speed_dist}%
Disturbance Scenario: Speed. The vertical light gray lines correspond to the time instants illustrated in Figs.~\ref{fig:nominal_dist}~and~\ref{fig:branching_dist}.}
\end{figure}

In the disturbance scenario, the other traffic participants do not behave as predicted. The Nominal MPC fails to perform a merging maneuver and the ego vehicle ends up in a crash with another vehicle as seen in Fig.~\ref{fig:nominal_dist}.
The ADSB-MPC strategy, however, can handle also this scenario, as can be observed in Fig.~\ref{fig:branching_dist}.
Like in the disturbance-free scenario, the ADSB-MPC in the disturbance scenario has an alternative plan to merge behind the white vehicle, if the white vehicle were to accelerate.
Because of this, the ADSB-MPC does not accelerate as heavily as the resulting control action from the Nominal MPC (see Fig.~\ref{fig:speed_dist}).
By 1.8\,s, where the Nominal MPC had reached an infeasible optimization problem, the ADSB-MPC is driving sufficiently slower than the white vehicle to plan a safe trajectory to merge behind it, and the ego vehicle eventually merges safely (see Fig.~\ref{fig:branching_dist}).

\subsection{Highway Driving}

\begin{figure}
\centering
\includegraphics{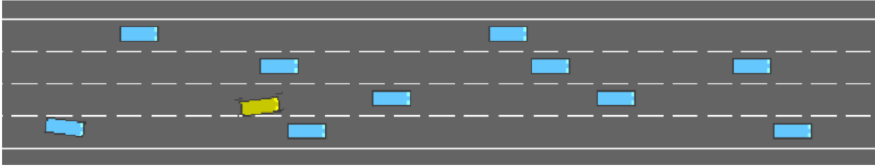}\\
\caption{\label{fig:highway-env}%
Capture from Highway-Env during execution of the ADSB-MPC planner.
The ego vehicle is shown as yellow.}
\end{figure}

To examine the controller performance in a tactical decision-making task, the ADSB-MPC is implemented for a highway driving scenario in the simulation environment Highway-Env \cite{highway-env} aimed at evaluating reinforcement-learning agents.
This enables comparison with reinforcement-learning methods and will test the ability of ADSB-MPC to handle a large number of randomized scenarios with, for the controller, unknown traffic behavior where multiple surrounding vehicles drive at varying velocities and overtake each other.
A number of reinforcement-learning agents were trained using Dueling
DQN \cite{10.5555/3045390.3045601}, QR-DQN \cite{Dabney_Rowland_Bellemare_Munos_2018}, PPO \cite{schulman2017proximal}, and TQC \cite{pmlr-v119-kuznetsov20a}, of which TQC performed the best and is used
here to compare with the ADSB-MPC planner.
A capture of a highway driving scenario in Highway-Env is shown in Fig.~\ref{fig:highway-env}.
The behavior of the traffic in Highway-Env is simulated using the Intelligent Driver Model (IDM) \cite{treiber} to determine the accelerations and the Minimizing Overall Braking Induced by Lane change (MOBIL) model \cite{doi:10.3141/1999-10} to determine lane changes.

The goal of the ego vehicle is to drive fast and keep to the rightmost lane without crashing into the other vehicles.
Specifically, the reward function used to evaluate the scenario and train the TQC agent is a linear combination of 
the reward obtained for crashing $r_\mathrm{crashed}$, 
the reward obtained for being closer to the correct lane $r_\mathrm{lane}$, 
and the reward for keeping the velocity high $r_\mathrm{speed}$:
\begin{equation} \label{eq:reward}
r = 
\frac{ b_c (1-r_\mathrm{crashed})
+ b_l r_\mathrm{lane}
+ b_v r_\mathrm{speed}}{b_c + b_l + b_v}
\end{equation}
where $b_c=1$, $b_l=0.1$, and $b_v=0.4$ are parameters. 
The crash reward $r_\mathrm{crashed}$ is 1 if the ego vehicle has crashed and 0 otherwise.
The lanes are numbered in increasing order from 0 to 3 from the leftmost lane to the rightmost lane.
The lane reward is determined by the current lane $i_\mathrm{lane} \in \{0,1,2,3\}$ and is given by
\begin{equation}
r_\mathrm{lane} = \frac{i_\mathrm{lane}}{n_\mathrm{lanes}-1}
\end{equation}
where $n_\mathrm{lanes}=4$ is the total number of lanes.
The speed reward $r_\mathrm{speed}$ is given by a linearly mapped function between 
$v_{r,\mathrm{min}}=20$\,m/s and $v_{r,\mathrm{max}}=30$\,m/s:
\begin{equation}
r_\mathrm{speed} = \begin{cases}
0, & v < v_{r,\mathrm{min}} \\
\frac{v - v_{r,\mathrm{min}}}{v_{r,\mathrm{max}} - v_{r,\mathrm{min}}}, & v_{r,\mathrm{min}} \leq v \leq v_{r,\mathrm{max}}. \\
1, & v > v_{r,\mathrm{max}}
\end{cases}
\end{equation}

The developed ADSB-MPC is compared to the nominal MPC and TQC in 100
episodes of 20\,s each, for three different traffic densities with all planners running at 5\,Hz.
  Here,
the pruning method proposed in
Sec.~\ref{sec:adversarial_objective_function} was used to only include
the $n=2$ most critical open-loop disturbance sequences in the
optimization problem for the ADSB-MPC.
The TQC agent was trained using the implementation from Stable-Baselines3 \cite{stable-baselines3}.
First, the TQC agent was trained for 5,000,000 steps in the default traffic density $d$,
then the TQC agent was trained for another 1,000,000 steps for each
step of increasingly denser traffic, first $1.5d$ and then $2d$,
resulting in three different TQC agents, TQC $d$, TQC $1.5d$, and TQC $2d$, specialized at driving in different
traffic densities.
Note that the objective of the comparison is not to find
the best performing reinforcement-learning approach, but to compare state-of-the-art
approaches to the proposed ADSB-MPC and discuss their properties.

Table~\ref{tab:highway-env}
shows the success rate as the number of episodes out of a 100 without
a collision and the accumulated reward as a percentage of the maximum
possible reward.
At low traffic density, success rate is high for
both MPC and TQC but TQC has a higher reward. It is important
to note that the MPC planners do not have the reward function \eqref{eq:reward} used in the evaluation
as an objective function, so this is not unexpected. When increasing
the traffic density by 50\,\%, the reward of the dedicated TQC planner
drops closer to that of the MPC planners, with a significantly lower success rate
than ADSB-MPC. Increasing the traffic density to double that of the
low traffic density, the ADSB-MPC outperforms the other planners,
including the TQC agent trained on high-density data. In summary, the
ADSB-MPC is the planner that is least likely to crash, which comes at
the cost of more careful driving than with the Nominal MPC.

\begin{table}[!t]
\renewcommand{\arraystretch}{1.3}
\caption{Success rates and accumulated rewards in Highway-Env}
\label{tab:highway-env}
\centering
\begin{tabular}{lccc}
\hline
\bfseries Traffic density & \bfseries Planner & \bfseries Success & \bfseries Reward \\
\hline
\multirow{3}{*}{Low / $d$}
& TQC $d$ & \underline{100/100} & \underline{99\,\%} \\
& Nominal MPC & 98/100 & 95\,\% \\
& ADSB-MPC & \underline{100/100} & 93\,\% \\
\hline
\multirow{3}{*}{Medium / $1.5d$}
& TQC $1.5d$ & 90/100 & \underline{93\,\%} \\
& Nominal MPC & 90/100 & 89\,\% \\
& ADSB-MPC & \underline{99/100} & 89\,\% \\
\hline
\multirow{3}{*}{High / $2d$}
& TQC $2d$ & 63/100 & 76\% \\
& Nominal MPC & 66/100 & 70\,\% \\
& ADSB-MPC & \underline{85/100} & \underline{77\,\%} \\
\hline
\end{tabular}
\end{table}

\section{Discussion}

A key aspect of ADSB-MPC is that the considered disturbance sequences are neither handcrafted nor is a statistical model of the disturbances needed. Instead, they are automatically determined as part of the ADSB-MPC strategy, taking the capabilities of the other agents in the scenario into account.
In the results in Sec.~\ref{sec:results}, the ADSB-MPC only used the simple prediction model that assumes the traffic to keep a constant velocity.
ADSB-MPC performs well in the randomized highway scenarios (see Table~\ref{tab:highway-env}) where the traffic is realistically modeled using IDM and MOBIL, indicating the applicability of the method on real-world driving.
As demonstrated in the merging scenario (see Fig.~\ref{fig:branching_dist}), unknown critical scenarios caused by other agents could be handled by ADSB-MPC if the possible actions of the agents are sufficiently modeled.
Being able to handle dangerous situations not previously encountered and that the system is not explicitly designed for is a promising property of ADSB-MPC.

Here, we studied a motion planner and controller with moderate control authority. For emergency maneuvers, where the problem is identified as infeasible with this level of control authority, a system with a higher control authority could be straightforwardly used.
Further, the method could be extended with a contingency scenario to, e.g., guarantee that it is possible to stop the ego vehicle before the current lane ends.

The ADSB-MPC approach results in an optimization-problem size that scales linearly with respect to the number of vehicles and the prediction horizon, which is appealing from a computational point-of-view.
ADSB-MPC also enables the problem size to be fixed with respect to the number of vehicles by using the adversarial objective function \eqref{eq:oppositional_objective} to only include the $n$ most critical disturbance sequences in the problem.
This is promising with regard to the scalability of the method when considering scenarios with many target vehicles.

Concerning the computational complexity, the procedure for determining the disturbance sequences is for the prototype implementation negligible compared to solving the optimal control problem in the MPC.
On average, approximately 5\,\% of the computational time is spent by the lattice planner to compute the disturbance sequences and the initial guess, 30\,\% of the time is spent in Ipopt to solve the optimization problem, 45\,\% of the time is spent computing expressions for derivatives to feed Ipopt before each sample, and the remainder is overhead spent in CasADi to formulate the optimal control problem and initialize Ipopt.
With few vehicles on the road, such as in the merging scenario, the prototype implementation runs in real-time on an i7-7700HQ CPU.

\section{Conclusions}

It is important that future driver-assistance systems and autonomous vehicles can handle hazardous events not encountered during their development, ADSB-MPC addresses the methodological development in this direction.
The proposed ADSB-MPC strategy uses automatically determined disturbance sequences that are intentionally adversarial to the ego vehicle.
To reduce conservatism, the concept of open-loop adversarial is used.
This leads to a resilient planning and control method, where the problem size scales linearly with the number of agents and the prediction horizon.
Further, the ADSB-MPC strategy can prune the problem to only include the most critical disturbance sequences to obtain a fixed problem size with respect to the number of agents.
The simulation results for multi-vehicle scenarios demonstrate that the ADSB-MPC can handle unknown critical scenarios without prior data of the potential hazard. 
Further, ADSB-MPC performs favorably to a TQC agent in high-density traffic without requiring prior training to obtain its policy and can negotiate traffic situations out of scope for a nominal MPC approach.


\bibliographystyle{IEEEtran}
\bibliography{IEEEabrv,ref}

\begin{thebibliography}{10}
\providecommand{\url}[1]{#1}
\csname url@samestyle\endcsname
\providecommand{\newblock}{\relax}
\providecommand{\bibinfo}[2]{#2}
\providecommand{\BIBentrySTDinterwordspacing}{\spaceskip=0pt\relax}
\providecommand{\BIBentryALTinterwordstretchfactor}{4}
\providecommand{\BIBentryALTinterwordspacing}{\spaceskip=\fontdimen2\font plus
\BIBentryALTinterwordstretchfactor\fontdimen3\font minus
  \fontdimen4\font\relax}
\providecommand{\BIBforeignlanguage}[2]{{%
\expandafter\ifx\csname l@#1\endcsname\relax
\typeout{** WARNING: IEEEtran.bst: No hyphenation pattern has been}%
\typeout{** loaded for the language `#1'. Using the pattern for}%
\typeout{** the default language instead.}%
\else
\language=\csname l@#1\endcsname
\fi
#2}}
\providecommand{\BIBdecl}{\relax}
\BIBdecl

\bibitem{schwall2020waymo}
\BIBentryALTinterwordspacing
M.~Schwall, T.~Daniel, T.~Victor, F.~Favaro, and H.~Hohnhold, ``Waymo public
  road safety performance data,'' 2020. [Online]. Available:
  \url{arXiv:2011.00038.}
\BIBentrySTDinterwordspacing

\bibitem{BOLN:2021:TITS}
B.~Olofsson and L.~Nielsen, ``Using crash databases to predict effectiveness of
  new autonomous vehicle maneuvers for lane-departure injury reduction,''
  \emph{{IEEE} Trans. Intell. Transp. Syst.}, vol.~22, no.~6, pp. 3479--3490,
  2021.

\bibitem{https://doi.org/10.1186/s40648-014-0001-z}
S.~Lef{\`{e}}vre, D.~Vasquez, and C.~Laugier, ``A survey on motion prediction
  and risk assessment for intelligent vehicles,'' \emph{Robomech J.}, vol.~1,
  no.~1, pp. 1--14, 2014.

\bibitem{9158529}
S.~Mozaffari, O.~Y. Al-Jarrah, M.~Dianati, P.~Jennings, and A.~Mouzakitis,
  ``Deep learning-based vehicle behavior prediction for autonomous driving
  applications: A review,'' \emph{{IEEE} Trans. Intell. Transp. Syst.},
  vol.~23, no.~1, pp. 33--47, 2022.

\bibitem{https://doi.org/10.1002/rob.21908}
O.~Ljungqvist, N.~Evestedt, D.~Axehill, M.~Cirillo, and H.~Pettersson, ``A path
  planning and path-following control framework for a general 2-trailer with a
  car-like tractor,'' \emph{J. Field Robot.}, vol.~36, no.~8, pp. 1345--1377,
  2019.

\bibitem{6145622}
C.~B. Browne, E.~Powley, D.~Whitehouse, S.~M. Lucas, P.~I. Cowling,
  P.~Rohlfshagen, S.~Tavener, D.~Perez, S.~Samothrakis, and S.~Colton, ``A
  survey of monte carlo tree search methods,'' \emph{{IEEE} Trans. Comput.
  Intell. {AI} in Games}, vol.~4, no.~1, pp. 1--43, 2012.

\bibitem{7535424}
D.~Lenz, T.~Kessler, and A.~Knoll, ``Tactical cooperative planning for
  autonomous highway driving using {M}onte-{C}arlo tree search,'' in \emph{2016
  IEEE Intell. Veh. Symp. (IV)}, 2016, pp. 447--453.

\bibitem{article}
V.~Mnih, K.~Kavukcuoglu, D.~Silver, A.~Rusu, J.~Veness, M.~Bellemare,
  A.~Graves, M.~Riedmiller, A.~Fidjeland, G.~Ostrovski, S.~Petersen,
  C.~Beattie, A.~Sadik, I.~Antonoglou, H.~King, D.~Kumaran, D.~Wierstra,
  S.~Legg, and D.~Hassabis, ``Human-level control through deep reinforcement
  learning,'' \emph{Nature}, vol. 518, pp. 529--33, 2015.

\bibitem{8569568}
C.-J. Hoel, K.~Wolff, and L.~Laine, ``Automated speed and lane change decision
  making using deep reinforcement learning,'' in \emph{21st Int. Conf. Intell.
  Transp. Syst. (ITSC)}, 2018, pp. 2148--2155.

\bibitem{alphago}
D.~Silver, A.~Huang, C.~Maddison, A.~Guez, L.~Sifre, G.~Driessche,
  J.~Schrittwieser, I.~Antonoglou, V.~Panneershelvam, M.~Lanctot, S.~Dieleman,
  D.~Grewe, J.~Nham, N.~Kalchbrenner, I.~Sutskever, T.~Lillicrap, M.~Leach,
  K.~Kavukcuoglu, T.~Graepel, and D.~Hassabis, ``Mastering the game of go with
  deep neural networks and tree search,'' \emph{Nature}, vol. 529, pp.
  484--489, 2016.

\bibitem{8911507}
C.-J. Hoel, K.~Driggs-Campbell, K.~Wolff, L.~Laine, and M.~J. Kochenderfer,
  ``Combining planning and deep reinforcement learning in tactical decision
  making for autonomous driving,'' \emph{{IEEE} Trans. Intell. Veh.}, vol.~5,
  no.~2, pp. 294--305, 2020.

\bibitem{10.5555/3045390.3045601}
Z.~Wang, T.~Schaul, M.~Hessel, H.~Van~Hasselt, M.~Lanctot, and N.~De~Freitas,
  ``Dueling network architectures for deep reinforcement learning,'' in
  \emph{Proc. 33rd Int. Conf. Mach. Lear. (ICML)}, vol.~48, New York, NY, 2016,
  pp. 1995--–2003.

\bibitem{Dabney_Rowland_Bellemare_Munos_2018}
W.~Dabney, M.~Rowland, M.~Bellemare, and R.~Munos, ``Distributional
  reinforcement learning with quantile regression,'' in \emph{Proc. 32nd AAAI
  Conf. AI}, vol.~32, no.~1, New Orleans, LA, 2018, pp. 2892--2901.

\bibitem{schulman2017proximal}
\BIBentryALTinterwordspacing
J.~Schulman, F.~Wolski, P.~Dhariwal, A.~Radford, and O.~Klimov, ``Proximal
  policy optimization algorithms,'' 2017. [Online]. Available:
  \url{arXiv:1707.06347.}
\BIBentrySTDinterwordspacing

\bibitem{pmlr-v119-kuznetsov20a}
A.~Kuznetsov, P.~Shvechikov, A.~Grishin, and D.~Vetrov, ``Controlling
  overestimation bias with truncated mixture of continuous distributional
  quantile critics,'' in \emph{Proc. 37th Int. Conf. Mach. Lear. (ICML)}, vol.
  119, Vienna, Austria, 2020, pp. 5556--5566.

\bibitem{Diehl}
M.~Diehl, H.~Bock, H.~Diedam, and P.-B. Wieber, \emph{Fast Direct Multiple
  Shooting Algorithms for Optimal Robot Control}.\hskip 1em plus 0.5em minus
  0.4em\relax Berlin, Heidelberg: Springer, 2006, vol. 340, pp. 65--93.

\bibitem{9084267}
K.~Bergman, O.~Ljungqvist, and D.~Axehill, ``Improved path planning by tightly
  combining lattice-based path planning and optimal control,'' \emph{{IEEE}
  Trans. Intell. Veh.}, vol.~6, no.~1, pp. 57--66, 2021.

\bibitem{9170864}
S.~Manzinger, C.~Pek, and M.~Althoff, ``Using reachable sets for trajectory
  planning of automated vehicles,'' \emph{{IEEE} Trans. Intell. Veh.}, vol.~6,
  no.~2, pp. 232--248, 2021.

\bibitem{9434948}
L.~Schäfer, S.~Manzinger, and M.~Althoff, ``Computation of solution spaces for
  optimization-based trajectory planning,'' \emph{{IEEE} Trans. Intell. Veh.},
  2021, to be published, doi: 10.1109/TIV.2021.3077702.

\bibitem{IV_Morsali_2021}
M.~Morsali, E.~Frisk, and J.~\AA{}slund, ``Spatio-temporal planning in
  multi-vehicle scenarios for autonomous vehicle using support vector
  machines,'' \emph{{IEEE} Trans. Intell. Veh.}, vol.~6, no.~4, pp. 611--621,
  2021.

\bibitem{MORARI1999667}
M.~Morari and J.~{H. Lee}, ``Model predictive control: past, present and
  future,'' \emph{Comput. Chem. Eng.}, vol.~23, no.~4, pp. 667--682, 1999.

\bibitem{KOTHARE19961361}
M.~V. Kothare, V.~Balakrishnan, and M.~Morari, ``Robust constrained model
  predictive control using linear matrix inequalities,'' \emph{Automatica},
  vol.~32, no.~10, pp. 1361--1379, 1996.

\bibitem{9410387}
T.~Brüdigam, M.~Olbrich, D.~Wollherr, and M.~Leibold, ``Stochastic model
  predictive control with a safety guarantee for automated driving,''
  \emph{{IEEE} Trans. Intell. Veh.}, to be published, doi:
  10.1109/TIV.2021.3074645.

\bibitem{LANGSON2004125}
W.~Langson, I.~Chryssochoos, S.~Raković, and D.~Mayne, ``Robust model
  predictive control using tubes,'' \emph{Automatica}, vol.~40, no.~1, pp.
  125--133, 2004.

\bibitem{lopez2019dynamic}
B.~T. Lopez, J.-J.~E. Slotine, and J.~P. How, ``Dynamic tube {MPC} for
  nonlinear systems,'' in \emph{2019 Amer. Control Conf. (ACC)}.\hskip 1em plus
  0.5em minus 0.4em\relax IEEE, 2019, pp. 1655--1662.

\bibitem{704989}
P.~Scokaert and D.~Mayne, ``Min-max feedback model predictive control for
  constrained linear systems,'' \emph{{IEEE} Trans. Autom. Control}, vol.~43,
  no.~8, pp. 1136--1142, 1998.

\bibitem{7990647}
G.~Cesari, G.~Schildbach, A.~Carvalho, and F.~Borrelli, ``Scenario model
  predictive control for lane change assistance and autonomous driving on
  highways,'' \emph{{IEEE} Intell. Transp. Syst. Mag.}, vol.~9, no.~3, pp.
  23--35, 2017.

\bibitem{9133136}
I.~Batkovic, U.~Rosolia, M.~Zanon, and P.~Falcone, ``A robust scenario {MPC}
  approach for uncertain multi-modal obstacles,'' \emph{{IEEE} Control Syst.
  Lett.}, vol.~5, no.~3, pp. 947--952, 2021.

\bibitem{6497657}
J.~Hardy and M.~Campbell, ``Contingency planning over probabilistic obstacle
  predictions for autonomous road vehicles,'' \emph{{IEEE} Trans. Robot.},
  vol.~29, no.~4, pp. 913--929, 2013.

\bibitem{alsterda2021contingency}
\BIBentryALTinterwordspacing
J.~P. Alsterda and J.~C. Gerdes, ``Contingency model predictive control for
  linear time-varying systems,'' 2021. [Online]. Available:
  \url{arXiv:2102.12045.}
\BIBentrySTDinterwordspacing

\bibitem{10.5555/1213331}
S.~M. LaValle, \emph{Planning Algorithms}.\hskip 1em plus 0.5em minus
  0.4em\relax Cambridge, U.K.: Cambridge Univ. Press, 2006.

\bibitem{ascher}
U.~M. Ascher and L.~R. Petzold, \emph{Computer methods for ordinary
  differential equations and differential-algebraic equations.}\hskip 1em plus
  0.5em minus 0.4em\relax Philadelphia, PA: SIAM, 1998.

\bibitem{casadi}
J.~A.~E. Andersson, J.~Gillis, G.~Horn, J.~B. Rawlings, and M.~Diehl,
  ``{CasADi} -- {A} software framework for nonlinear optimization and optimal
  control,'' \emph{Math. Program. Comput.}, vol.~11, no.~1, pp. 1--36, 2019.

\bibitem{ipopt}
A.~W{\"a}chter and L.~T. Biegler, ``On the implementation of an interior-point
  filter line-search algorithm for large-scale nonlinear programming,''
  \emph{Math. Program.}, vol. 106, no.~1, pp. 25--57, 2006.

\bibitem{hsl}
\BIBentryALTinterwordspacing
{HSL}, ``{A} collection of {F}ortran codes for large scale scientific
  computation,'' 2021, {A}ccessed: 2021-09-08. [Online]. Available:
  \url{http://www.hsl.rl.ac.uk}
\BIBentrySTDinterwordspacing

\bibitem{franklin2009feedback}
G.~F. Franklin, J.~D. Powell, A.~Emami-Naeini, and J.~D. Powell, \emph{Feedback
  control of dynamic systems}, 6th~ed.\hskip 1em plus 0.5em minus 0.4em\relax
  Upper Saddle River, NJ: Prentice-Hall, 2009.

\bibitem{highway-env}
E.~Leurent, ``An environment for autonomous driving decision-making,''
  \url{https://github.com/eleurent/highway-env}, 2018, {A}ccessed: 2022-02-18.

\bibitem{treiber}
M.~Treiber, A.~Hennecke, and D.~Helbing, ``Congested traffic states in
  empirical observations and microscopic simulations,'' \emph{Phys. Rev. E:
  Stat. Phys., Plasmas, Fluids, Relat. Interdiscip. Top.}, vol.~62, pp.
  1805--24, 2000.

\bibitem{doi:10.3141/1999-10}
A.~Kesting, M.~Treiber, and D.~Helbing, ``General lane-changing model {MOBIL}
  for car-following models,'' \emph{Transp. Res. Rec.}, vol. 1999, no.~1, pp.
  86--94, 2007.

\bibitem{stable-baselines3}
A.~Raffin, A.~Hill, A.~Gleave, A.~Kanervisto, M.~Ernestus, and N.~Dormann,
  ``Stable-baselines3: Reliable reinforcement learning implementations,''
  \emph{J. Mach. Lear. Res.}, vol.~22, no. 268, pp. 1--8, 2021.

\end{thebibliography}

\newpage

\begin{IEEEbiography}[{\includegraphics[width=1in,height=1.25in,clip,keepaspectratio,trim={0.4cm 0 0.4cm 0}]{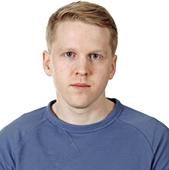}}]
{Victor Fors} received the M.Sc. degree in Engineering Physics in 2015 from Lund University, Lund, Sweden
and his Ph.D. degree in electrical engineering in 2020 from Link{\"o}ping University, Link{\"o}ping, Sweden.
He is currently a postdoc at Stanford University, Stanford, CA, USA.
His main research interests are in vehicle dynamics, control, active safety, and autonomy.
\end{IEEEbiography}%
\begin{IEEEbiography}[{\includegraphics[width=1in,height=1.25in,clip,keepaspectratio,trim={0.4cm 0 0.4cm 0}]{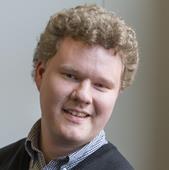}}]
{Bj{\"o}rn Olofsson} received the M.Sc. degree in Engineering Physics in 2010 and the Ph.D. degree in Automatic Control in 2015, both from Lund University, Sweden. He is currently a researcher at the Division of Vehicular Systems, Link{\"o}ping University, Link{\"o}ping, Sweden and at the Department of Automatic Control, Lund University, Lund, Sweden. His research interests are in motion control for robots and vehicles, optimal control, system identification, and statistical sensor fusion.
\end{IEEEbiography}%
\begin{IEEEbiography}[{\includegraphics[width=1in,height=1.25in,clip,keepaspectratio,trim={0.4cm 0 0.4cm 0}]{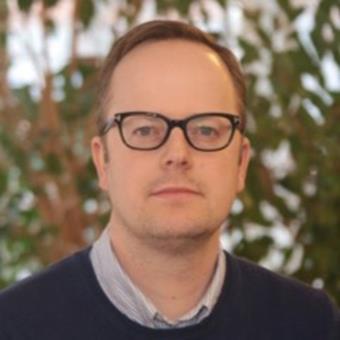}}]
  {Erik Frisk} was born in Stockholm, Sweden, in 1971. He received the
  Ph.D. degree in 2001. He is currently a Professor with the
  Department of Electrical Engineering, Linköping University,
  Linköping, Sweden. His main research interests are optimization
  techniques for autonomous vehicles in complex traffic scenarios and
  model and data-driven fault diagnostics and prognostics.
\end{IEEEbiography}

\vfill

\end{document}